\documentclass[aps,prd,twocolumn,superscriptaddress,showpacs,amsmath,amssymb]{revtex4}
\usepackage{graphicx,epsf,amsmath}
\usepackage[]{latexsym}
\newcommand{\be}{\begin{eqnarray}}
\newcommand{\ee}{\end{eqnarray}}

\newcommand{\ket}[1]{\mbox{$\mid #1\,\rangle$}}

\newcommand{\pro}[2]{\mbox{$\langle\, #1 \mid #2\,\rangle$}}
\newcommand{\expec}[1]{\mbox{$\langle\, #1\,\rangle$}}

\renewcommand{\d}{\mbox{${\rm d}$}}
\newcommand{\lp}{\ell_{\rm p}}
\newcommand{\mpl}{m_{\rm p}}

\newcommand{\rh}{R_{\rm H}}
\begin{document}
\title{Horizon wave-function for single localized particles:
\\
GUP and quantum black hole decay}
\author{Roberto~Casadio}
\email{casadio@bo.infn.it}
\affiliation{Dipartimento di Fisica e Astronomia, Universit\`a di Bologna,
via~Irnerio~46, 40126~Bologna, Italy}
\affiliation{I.N.F.N., Sezione di Bologna, viale Berti~Pichat~6/2, 40127~Bologna, Italy}
\author{Fabio Scardigli}
\email{fabio@phys.ntu.edu.tw}
\affiliation{Dipartimento di Matematica, Politecnico di Milano, Piazza L.~da Vinci~32,
20133~Milano, Italy}
\affiliation{Yukawa Institute for Theoretical Physics,
Kyoto University, Kyoto 606-8502, Japan}
\begin{abstract}
A localised particle in Quantum Mechanics is described by a wave packet
in position space, regardless of its energy.
However, from the point of view of General Relativity, if the particle's energy density
exceeds a certain threshold, it should be a black hole.
In order to combine these two pictures, we introduce a horizon wave-function determined
by the particle wave-function in position space, which eventually yields the probability that
the particle is a black hole.
The existence of a minimum mass for black holes naturally follows,
albeit not in the form of a sharp value around the Planck scale, but rather like a vanishing
probability that a particle much lighter than the Planck mass be a black hole.
We also show that our construction entails an effective Generalised Uncertainty Principle
(GUP), simply obtained by adding the uncertainties coming from the two wave-functions
associated to a particle.
Finally, the decay of microscopic (quantum) black holes is also described in agreement
with what the GUP predicts.
\end{abstract}
\maketitle
\section{Introduction and motivation}
Understanding all the physical aspects in the gravitational collapse of a compact object
and how black holes form, remains one of the most intriguing challenges of contemporary
theoretical physics.
After the seminal papers of Oppenheimer and co-workers~\cite{OS}, the literature
on the subject has grown immensely, but many issues are still open in General Relativity
(see, e.g., Refs.~\cite{joshi,Bekenstein:2004eh}, and references therein), not to mention
the conceptual and technical difficulties one faces when the quantum nature of the
collapsing matter is taken into account.
Assuming quantum gravitational fluctuations are small, one can describe matter
by means of Quantum Field Theory on the curved background space-time, an
approach which has produced remarkable results, but is unlikely to be directly
applicable to a self-gravitating system representing a collapsing object.
\par
A general property of the Einstein theory is that the gravitational interaction is always
attractive and we are thus not allowed to neglect its effect on the causal structure
of space-time if we pack enough energy in a sufficiently small volume.
This can occur, for example, if two particles (for simplicity, of negligible spatial extension and
total angular momentum) collide with an impact parameter $b$ shorter than the Schwarzschild
radius corresponding to the total center-mass energy $E$ of the system,
that is~\footnote{We shall use units with $c=k_B=1$,
and always display the Newton constant $G=\lp/\mpl$, where $\lp$ and $\mpl$
are the Planck length and mass, respectively, so that $\hbar=\lp\,\mpl$.}
\be
b\lesssim 2\,\lp\,\frac{E}{\mpl}
\equiv
\rh
\ .
\label{hoop}
\ee
This {\em hoop conjecture\/}~\cite{Thorne:1972ji} has been checked and verified
theoretically in a variety of situations, but it was initially formulated for black holes
of (at least) astrophysical size~\cite{payne}, for which the very concept of a classical
background metric and related horizon structure should be reasonably safe
(for a review of some problems, see the bibliography in Ref.~\cite{Senovilla:2007dw}).
Whether the concepts involved in the above conclusion can also be trusted for
masses approaching the Planck size, however, is definitely more challenging.
In fact, for masses in that range, quantum effects may hardly be neglected
(for a recent discussion, see, e.g., Ref.~\cite{acmo})
and it is reasonable that the picture arising from General Relativistic black holes must
be replaced in order to include the possible existence of new objects,
generically referred to as ``quantum black holes'' (see, e.g., Refs.~\cite{hsu,calmet}).
\par
The main complication in studying the Planck regime is that we do not have any
experimental insight thereof, which makes it very difficult to tell whether any theory
we could come up with is physically relevant.
We might instead start from our established concepts and knowledge of nature,
and push them beyond the present experimental limits.
If we set out to do so, we immediately meet with a conceptual challenge:
how can we describe a system containing both Quantum Mechanical objects
(such as the elementary particles of the Standard Model) and classically defined
horizons?
The aim of this paper is precisely to introduce the definition of a wave
function for the horizon that can be associated with any localised Quantum
Mechanical particle~\cite{Cfuzzy}.
This tool will allow us to put on quantitative ground the condition that
distinguishes a black hole from a regular particle.
And we shall also see that our construction naturally leads to an effective
Generalised Uncertainty Principle (GUP)~\cite{scardigli} for the particle position,
and a decay rate for microscopic black holes.
\par
The paper is organised as follows:
in the next Section we introduce the main ideas that define the horizon wave-function
associated with any localised Quantum Mechanical particle;
in Section~\ref{Gparticle}, we then apply the general construction to the particularly simple
case of a particle described by a Gaussian wave-function at rest in flat space-time,
for which we explicitly obtain the probability that the particle is a black hole,
we recover the GUP and a minimum measurable length, and estimate the decay
rate of a black hole with mass around the Planck scale;
finally, in Section~\ref{conc}, we comment on our findings and outline future
applications.
\section{Horizon Quantum Mechanics}
Given a matter source, say a spherically symmetric ``particle'',
General Relativity and Quantum Mechanics naturally associate with it two length
scales:
the Schwarzschild radius and the Compton-de~Broglie wavelength, respectively.
We shall therefore start by briefly reviewing these concepts and then propose how
to extend the former into the realm of Quantum Mechanics, where the latter is born.
\subsection{Spherical trapping horizons}
The appearance of a classical horizon is relatively easy to understand
in a spherically symmetric space-time.
Let us first recall that we can write a general spherically symmetric metric $g_{\mu\nu}$
as
\be
\d s^2
=
g_{ij}\,\d x^i\,\d x^j
+
r^2(x^i)\left(\d\theta^2+\sin^2\theta\,\d\phi^2\right)
\ ,
\label{metric}
\ee
where $r$ is the areal coordinate and $x^i=(x^1,x^2)$ are coordinates
on surfaces where the angles $\theta$ and $\phi$ are constant.
The location of a trapping horizon, a surface where the escape velocity equals
the speed of light~\footnote{More technically, a trapping surface is the
location where the divergence of outgoing null congruences vanishes.},
is then determined by the equation~\cite{hayward}
\be
0
=
g^{ij}\,\nabla_i r\,\nabla_j r
=
1-\frac{2\,M}{r}
\ ,
\label{th}
\ee
where $\nabla_i r$ is the covector perpendicular to surfaces of constant area
$\mathcal{A}=4\,\pi\,r^2$.
The function $M=\lp\,m/\mpl$ is the active gravitational (or Misner-Sharp)
mass, representing the total energy enclosed within a sphere of radius $r$.
For example, if we set $x^1=t$ and $x^2=r$, the function $m$ is explicitly given
by the integral of the classical matter density $\rho=\rho(x^i)$
weighted by the flat metric volume measure,
\be
m(t,r)=\frac{4\,\pi}{3}\int_0^r \rho(t, \bar r)\,\bar r^2\,\d \bar r
\ ,
\label{M}
\ee
as if the space inside the sphere were flat.
Of course, it is in general very difficult to follow the dynamics of a given
matter distribution and verify the existence of surfaces satisfying Eq.~\eqref{th},
but we can say an horizon exists if there are values of $r$ such that
\be
R_M=2\,M(t,r)
\ge
r
\ ,
\ee
which generalises the hoop conjecture~\eqref{hoop} to continuous energy densities
(in fact, the horizon radius saturates the above inequality, i.e.~$\rh=r$).
\par
Note the above equation does not lead to any mass threshold for the existence of
a black hole, since $M$ is not limited from below in the classical theory, and the
area of the trapping surface can be vanishingly small.
However, if we consider a spin-less point-like source of mass $m$,
Quantum Mechanics introduces an uncertainty in its spatial localisation,
typically of the order of the Compton length,
\be
\lambda_m
\simeq
\lp\,\frac{\mpl}{m}
=
\frac{\lp^2}{M}
\ .
\label{lambdaM}
\ee
Assuming quantum physics is a more refined description of classical physics,
the clash of the two lengths, $\rh$ and $\lambda_m$, implies that the former
only makes sense provided it is larger than the latter,
\be
\rh
\gtrsim
\lambda_m
\quad
\Rightarrow
\quad
m
\gtrsim
\mpl
\ ,
\label{clM}
\ee
or $M\gtrsim\lp$.
Note that this argument employs the flat space Compton length~\eqref{lambdaM},
and it is likely that the particle's self-gravity will affect it.
However, it is still reasonable to assume the condition~\eqref{clM} holds as an
order of magnitude estimate.
\par
Overall, the common argument that quantum gravity effects should
become relevant only at scales of order $\mpl$ or higher now appears questionable,
since the condition~\eqref{clM} implies that such a system can be fairly well-described
in classical terms.
This is indeed at the core of the idea of ``classicalization'' given in Ref.~\cite{dvali}
and, before that, of gravitationally inspired GUPs~\cite{scardigli,BNSminimallength}.
In particular, following the canonical steps that lead to the construction of Quantum
Mechanics, the latter are usually assumed to hold as fundamental principles for the
reformulation of Quantum Mechanics in the presence of gravity.
Note then that gravity would reduce to a ``kinematical effect'' encoded by the modified
commutators for the canonical variables.
In the following we shall instead start from the introduction of a auxiliary wave-function
that describes the horizon associated with a given localised particle, and show that
a modified uncertainty relation follows consequently.
\subsection{Horizon Wave-function}
\label{Hwf}
Let us first formulate the construction in a somewhat general fashion.
For simplicity, we shall only consider quantum mechanical states representing
{\em spherically symmetric\/} objects, which are both {\em localised in space\/}
and {\em at rest\/} in the chosen reference frame.
The particle is consequently described by a wave-function
$\psi_{\rm S}\in L^2(\mathbb{R}^3)$,
which we assume can be decomposed into energy eigenstates,
\be
\ket{\psi_{\rm S}}
=
\sum_E\,C(E)\,\ket{\psi_E}
\ ,
\ee
where the sum represents the spectral decomposition in Hamiltonian eigenmodes,
\be
\hat H\,\ket{\psi_E}=E\,\ket{\psi_E}
\ ,
\ee
and the actual Hamiltonian $H$ needs not be specified yet~\footnote{This is where, for instance,
the self-gravity of the particle may enter.}.
The expression of the Schwarzschild radius in Eq.~\eqref{hoop} can be inverted to obtain
\be
E
=
\mpl\,\frac{\rh}{2\,\lp}
\ ,
\ee
and we then define the (unnormalised) ``horizon wave-function'' as
$\tilde\psi_{\rm H}(\rh)=C\left(\mpl\,{\rh}/{2\,\lp}\right)$,
whose normalisation is fixed by assuming the scalar product
\be
\pro{\psi_{\rm H}}{\phi_{\rm H}}
=
4\,\pi\,\int_0^\infty
\psi_{\rm H}^*(\rh)\,\phi_{\rm H}(\rh)\,\rh^2\,\d \rh
\ .
\label{normH}
\ee
We could now simply say that the normalised wave-function $\psi_{\rm H}$ yields
the probability that an observer would detect a horizon of areal radius $r=\rh$
associated with the particle in the quantum state $\psi_{\rm S}$.
Such a horizon would necessarily be ``fuzzy'', like is the position of the particle itself,
but giving such a claim an experimental meaning does not appear very simple.
\par
A more precise use of the horizon wave-function can however be already outlined.
For example, having defined the wave-function $\psi_{\rm H}$ associated with
a given $\psi_{\rm S}$, the probability density that the particle lies inside its own horizon
of radius $r=\rh$ will be given by
\be
P_<(r<\rh)
=
P_{\rm S}(r<\rh)\,P_{\rm H}(\rh)
\ ,
\label{PrlessH}
\ee
where
\be
P_{\rm S}(r<\rh)
=
4\,\pi\,\int_0^{\rh}
|\psi_{\rm S}(r)|^2\,r^2\,\d r
\ee
is the probability that the particle is inside a sphere of radius $r=\rh$,
and
\be
P_{\rm H}(\rh)
=
4\,\pi\,\rh^2\,|\psi_{\rm H}(\rh)|^2
\label{Ph}
\ee
is the probability that the sphere of radius $r=\rh$ is a horizon.
Finally, the probability that the particle described by the wave-function $\psi_{\rm S}$ is a
black hole will be obtained by integrating~\eqref{PrlessH} over all possible
values of the horizon radius, namely
\be
P_{\rm BH}
=
\int_0^\infty P_<(r<\rh)\,\d \rh
\ .
\label{PBH}
\ee
It is this final probability we now proceed to clarify with an example, along with
a derivation of a GUP and some predictions for the decay of a quantum black hole.
\section{Gaussian packet at rest in flat space}
\label{Gparticle}
Assuming the space-time is flat, our construction can be exemplified by describing
the massive particle at rest in the origin of the reference frame with the spherically
symmetric Gaussian wave-function
\be
\psi_{\rm S}(r)
=
\frac{e^{-\frac{r^2}{2\,\ell^2}}}{\ell^{3/2}\,\pi^{3/4}}
\ ,
\ee
where we shall usually assume that the width $\ell$ is given by the Compton
length~\eqref{lambdaM} of the particle,
\be
\ell
=
\lambda_m
\simeq
\lp\,\frac{\mpl}{m}
\ .
\ee
The above packet corresponds to the momentum space wave-function
\be
\psi_{\rm S}(p)
=
\frac{e^{-\frac{p^2}{2\,\Delta^2}}}{\Delta^{3/2}\,\pi^{3/4}}\,
\ ,
\ee
where $p^2=\vec p\cdot\vec p$ is the square modulus of the spatial momentum,
and the width
\be
\Delta
=
\mpl\,\frac{\lp}{\ell}
\simeq
m
\ .
\ee
For the energy of the particle, we can simply assume the relativistic mass-shell relation
in flat space,
\be
E^2=p^2+m^2
\ ,
\ee
and, upon inverting the expression of the Schwarzschild radius~\eqref{hoop},
we obtain the unnormalized wave-function
\be
\tilde \psi_H(\rh)
=
\frac{\ell^{3/2}\,e^{\frac{\ell^2\,m^2}{2\,\lp^2\,\mpl^2}}\,
e^{-\frac{\lp^2\,\rh^2}{8\,\lp^4}}}
{\pi^{3/4}\,\lp^{3/2}\,\mpl^{3/2}}
\ .
\ee
Finally, the inner product~\eqref{normH} yields the normalized horizon
wave-function
\be
\psi_H(\rh)
=
\frac{\ell^{3/2}\,e^{-\frac{\ell^2\,\rh^2}{8\,\lp^4}}}
{2^{3/2}\,\pi^{3/4}\,\lp^3}
\ .
\ee
\begin{figure}[t]
\centering
\raisebox{3.5cm}{$P$}
\includegraphics[width=7cm]{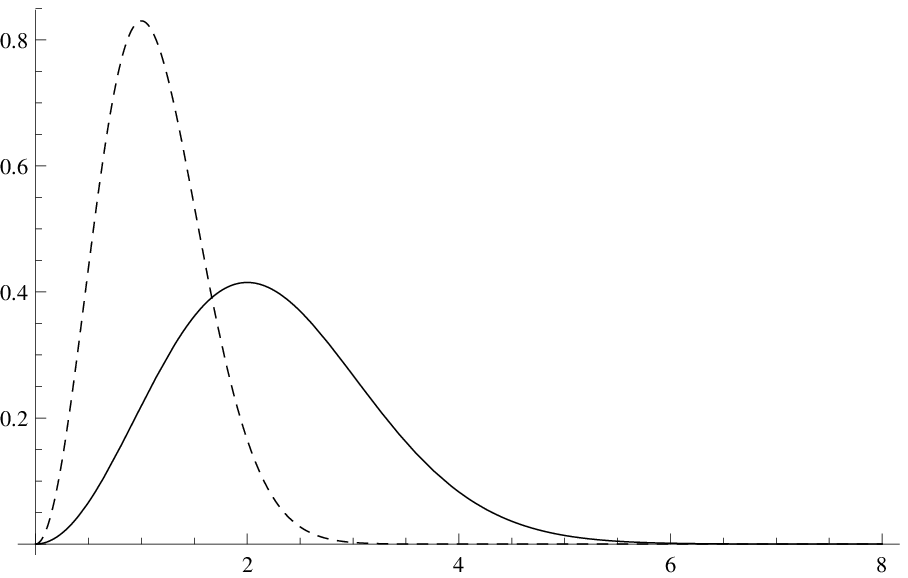}
\\
\hspace{7cm}$r/\lp$
\\
\raisebox{3.5cm}{$P$}
\includegraphics[width=7cm]{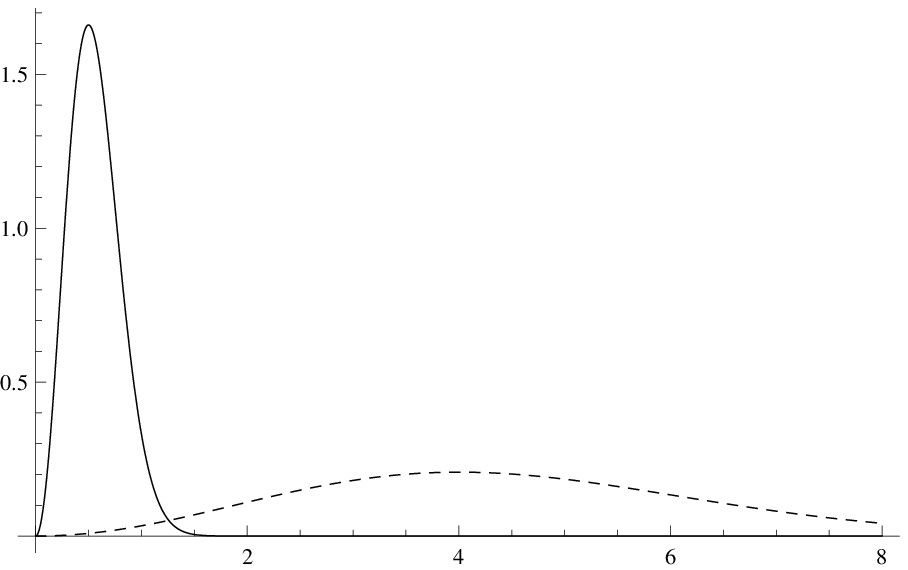}
\\
\hspace{7cm}$r/\lp$
\caption{Probabilities $P_{\rm H}$ in Eq.~\eqref{Ph} (dashed line) and $P_{\rm S}$
in Eq.~\eqref{Ps} (solid line)
for $m=\mpl/2$ (upper panel) and $m=2\,\mpl$ (lower panel), assuming $m\sim \ell^{-1}$.
\label{2Psi}}
\end{figure}
\begin{figure}[t]
\centering
\raisebox{3.5cm}{$P_<$}
\includegraphics[width=7cm]{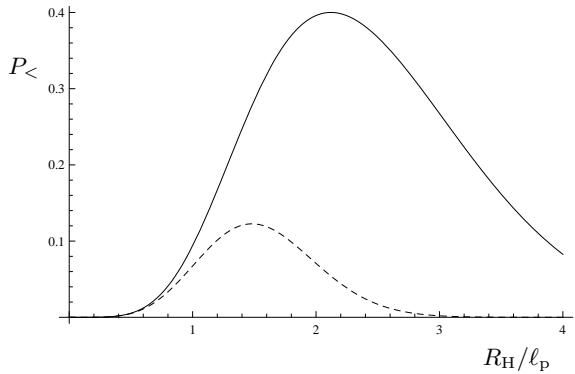}
\\
\hspace{6cm}$\rh/\lp$
\caption{Probability density $P_<$ in Eq.~\eqref{Pin} that particle is inside its horizon of radius $R=\rh$,
for $\ell=\lp$ (solid line) and for $\ell=2\,\lp$ (dashed line).
\label{prob<}}
\end{figure}
\begin{figure}[t]
\centering
\raisebox{3.5cm}{$P_{\rm BH}$}
\includegraphics[width=7cm]{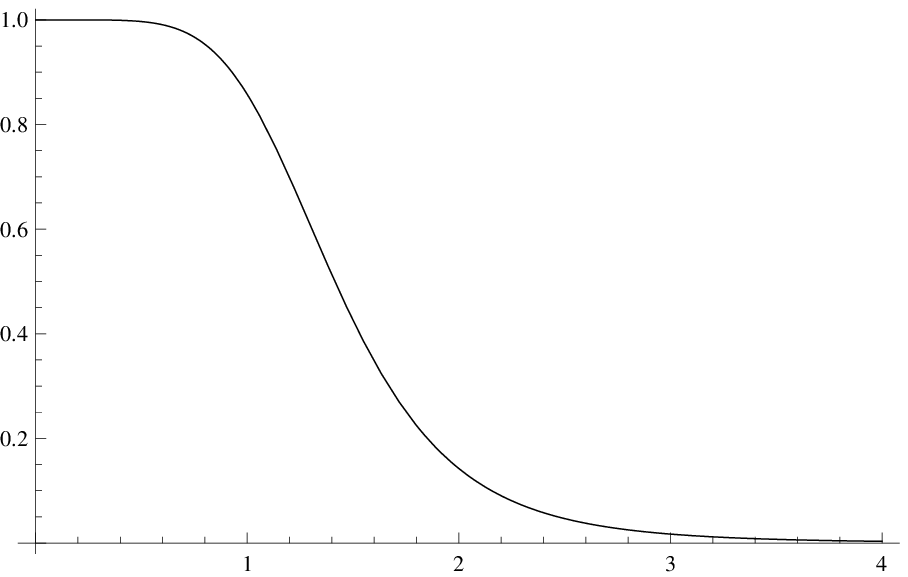}
\\
\hspace{7cm}$\ell/\lp$
\\
\raisebox{3.5cm}{$P_{\rm BH}$}
\includegraphics[width=7cm]{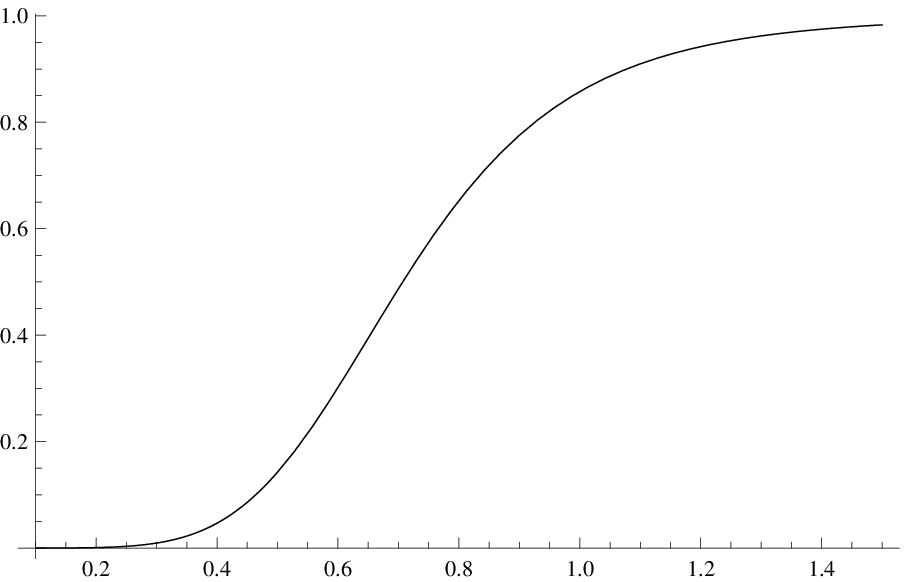}
\\
\hspace{7cm}$m/\mpl$
\caption{Probability $P_{\rm BH}$ in Eq.~\eqref{Pbh} that particle of width $\ell\sim m^{-1}$ is a black hole.
\label{prob}}
\end{figure}
\par
Note that, since
$\expec{\hat r^2}\simeq \ell^2$ and $\expec{\hat R_{\rm H}^2}\simeq \lp^4/\ell^2$,
we expect the particle will be inside its own horizon if $\expec{\hat r^2}\ll \expec{\hat R_{\rm H}^2}$,
which precisely yields the condition~\eqref{clM} if $\ell\sim m^{-1}$.
This is clear, for example, in Fig.~\ref{2Psi}, where the probability $P_{\rm H}=P_{\rm H}(r)$ is plotted
along with the probability
\be
P_{\rm S}(r)
=
4\,\pi\,r^2\,|\psi_{\rm S}(r)|^2
\ ,
\label{Ps}
\ee
for $m\lesssim\mpl$ and $m\gtrsim\mpl$.
In the former case, the horizon is more likely found with a smaller radius than
the particle's, with the opposite occurring in the latter.
In fact, the probability density~\eqref{PrlessH} can now be explicitly computed,
\be
P_<
=
\frac{\ell^3\,\rh^2\,e^{-\frac{\ell^2\,\rh^2}{4\,\lp^4}}}{2\,\sqrt{\pi}\,\lp^6}
\left[
{\rm Erf}\left(\frac{\rh}{\ell}\right)
-
\frac{2\,\rh\,e^{-\frac{\rh^2}{\ell^2}}}{\sqrt{\pi}\,\ell}
\right]
\ ,
\label{Pin}
\ee
from which the probability~\eqref{PBH} for the particle to be a black hole is obtained as
\be
P_{\rm BH}(\ell)
=
\frac{2}{\pi}\left[
\arctan\left(2\,\frac{\lp^2}{\ell^2}\right)
+
2\,\frac{\ell^2\,(4-\ell^4/\lp^4)}{\lp^2\,(4+\ell^4/\lp^4)^2}
\right]
\ ,
\label{Pbh}
\ee
or, writing $P_{\rm BH}$ as a function of $m$,
\be
P_{\rm BH}
=
\frac{2}{\pi}\left[
\arctan\left(2\,\frac{m^2}{\mpl^2}\right)
+
2\,\frac{\mpl^2\,(4-\mpl^4/m^4)}{m^2\,(4+\mpl^4/m^4)^2}
\right]
\ .
\label{Pbh2}
\ee
In Fig.~\ref{prob<}, we show the probability density~\eqref{Pin}, for two different
values of the Gaussian width $\ell$.
Since $\ell\sim m^{-1}$, it is already clear that such probability decreases for decreasing
$m$ (below the Planck mass).
In fact, in Fig.~\ref{prob}, we show the probability~\eqref{Pbh} that the particle is a black hole
as a function of the Gaussian width $\ell$ (upper panel) and particle mass $m\sim\ell^{-1}$
(lower panel).
From the plot of $P_{\rm BH}$, it appears pretty obvious that the particle is most likely a black hole,
$P_{\rm BH}\simeq 1$, if $\ell\lesssim\lp$.
Assuming as usual $\ell\sim m^{-1}$, we have thus derived the same condition~\eqref{clM},
from a totally Quantum Mechanical picture.
\par
An important remark is that we have here assumed flat space throughout the computation,
which means the self-gravity of the particle has been neglected.
It is very likely that such an approximation fails for large black holes with $m\gg\mpl$,
although the general idea outlined in Section~\ref{Hwf} should still be valid.
Of course, one could then improve the description of particles with $m\gg\mpl$ by
employing a curved-space mass-shell relation and suitable normal modes, rather
than simple plane waves.
\subsection{Effective GUP}
\begin{figure}[t]
\centering
\raisebox{3.5cm}{$\frac{\Delta r}{\lp}$}
\includegraphics[width=8cm]{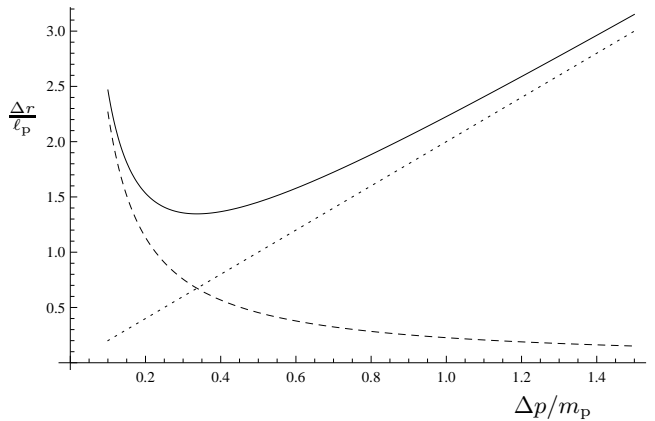}
\\
\hspace{6cm}$\Delta p/\mpl$
\caption{Uncertainty relation~\eqref{effGUP} (solid line) as a combination
of the Quantum Mechanical uncertainty (dashed line) and the uncertainty
in horizon radius (dotted line).
\label{pGUP}}
\end{figure}

For the Gaussian packet described above, it is easy to find that the usual Quantum
Mechanical uncertainty in radial position is given by
\be
\expec{\Delta r^2}
&\!\!=\!\!&
4\,\pi\,\int_0^{\infty}
|\psi_{\rm S}(r)|^2\,r^4\,\d r
\nonumber
\\
&&
-
\left(
4\,\pi\,\int_0^{\infty}
|\psi_{\rm S}(r)|^2\,r^3\,\d r
\right)^2
\nonumber
\\
&\!\!=\!\!&
\left(\frac{3\,\pi-8}{2\,\pi}\right)
\ell^2
\ .
\label{Dr}
\ee
Analogously, the uncertainty in the horizon radius will be given by
\be
\expec{\Delta \rh^2}
&\!\!=\!\!&
4\,\pi\,\int_0^{\infty}
|\psi_{\rm H}(\rh)|^2\,\rh^4\,\d \rh
\nonumber
\\
&&
-
\left(
4\,\pi\,\int_0^{\infty}
|\psi_{\rm H}(\rh)|^2\,\rh^3\,\d \rh
\right)^2
\nonumber
\\
&\!\!=\!\!&
4\left(\frac{3\,\pi-8}{2\,\pi}\right)
\frac{\lp^4}{\ell^2}
\ .
\label{DRH}
\ee
Since
\be
\expec{\Delta p^2}
&\!\!=\!\!&
4\,\pi\,\int_0^{\infty}
|\psi_{\rm S}(p)|^2\,p^4\,\d p
\nonumber
\\
&&
-
\left(
4\,\pi\,\int_0^{\infty}
|\psi_{\rm S}(p)|^2\,p^3\,\d p
\right)^2
\nonumber
\\
&\!\!=\!\!&
\left(\frac{3\,\pi-8}{2\,\pi}\right)
\mpl^2\,\frac{\lp^2}{\ell^2}
\equiv
\Delta p^2
\ ,
\ee
we can also write
\be
\ell^2
=
\left(\frac{3\,\pi-8}{2\,\pi}\right)
\lp^2\,\frac{\mpl^2}{\Delta p^2}
\ .
\ee
Finally, by combining the uncertainty~\eqref{Dr} with \eqref{DRH} linearly,
we find
\be
\Delta r
&\!\!\equiv\!\!&
\sqrt{\expec{\Delta r^2}}
+
\gamma\,
\sqrt{\expec{\Delta \rh^2}}
\nonumber
\\
&\!\!=\!\!&
\left(\frac{3\,\pi-8}{2\,\pi}\right)
\lp\,\frac{\mpl}{\Delta p}
+
2\,\gamma\,\lp\,\frac{\Delta p}{\mpl}
\ ,
\label{effGUP}
\ee
where $\gamma$ is a coefficient of order one, and the result is
plotted in Fig.~\ref{pGUP} (for $\gamma=1$).
This is precisely the kind of GUP considered in Refs.~\cite{scardigli},
leading to a minimum measurable length
\be
\Delta r
\ge
2\,\sqrt{\gamma\,\frac{3\,\pi-8}{\pi}}\,\lp
\simeq
1.3\,\sqrt{\gamma}\,\lp
\ ,
\ee
obtained for
\be
\Delta p=\sqrt{\frac{3\,\pi-8}{\pi\,\gamma}}\,\frac{\mpl}{2}
\ .
\ee
\par
Of course, one might consider different ways of combining the two
uncertainties~\eqref{Dr} and \eqref{DRH}, or even avoid this step and
just make a direct use of the horizon wave-function.
In this respect, the present approach appears more flexible, provided
one is able to extend it to different physical systems, as we shall
further discuss in the last Section.
\subsection{Quantum black hole evaporation}
\begin{figure}[t]
\centering
\raisebox{3.5cm}{$\frac{T}{\mpl}$}
\includegraphics[width=8cm]{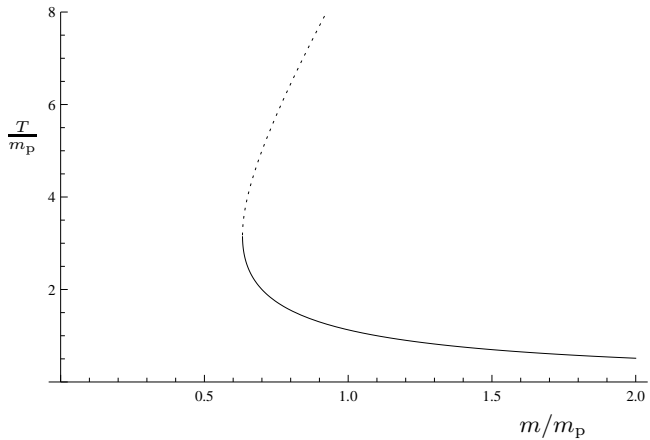}
\\
\hspace{6cm}$m/\mpl$
\caption{Temperature vs.~mass according to Eq.~\eqref{mT}
with $\beta=1/10$:
solid line reproduces the Hawking behaviour for large $m\gg\mpl$;
dotted line is the unphysical branch, and their meeting point
represents the black hole with minimum mass.
\label{mTp}}
\end{figure}

The well-known result due to Hawking~\cite{hawking},
\be
T_{\rm H}
=
\frac{\mpl^2}{8\,\pi\,m}
\ ,
\ee
extrapolated to vanishingly small mass $M$ implies that $T_{\rm H}$
diverges.
On the other hand, one can derive modified black hole temperatures
for $m\simeq\mpl$ from the GUP~\cite{fabio,WC}.
In particular, we just recall that one obtains
\be
m
=
\frac{\mpl^2}{8\,\pi\, T}
+2\,\pi\,\beta\,T
\ ,
\label{mT}
\ee
where
\be
\beta = \frac{\gamma}{4\,\pi\,(3\,\pi -8)}
>0
\ ,
\ee
in order to ensure the existence of a minimum mass
for the black hole (see Fig.~\ref{mTp}).
This is a consistency condition with the result that $P_{\rm BH}\simeq 1$
only for $m\gtrsim \mpl$, or that one does not have a black hole
for masses significantly smaller then $\mpl$.
In fact, from (\ref{mT}) we get
\be
m_{\rm{min}}
=
\sqrt{\beta}\,\mpl
\ ,
\qquad
T_{\rm{max}}
=
\frac{\mpl}{4\,\pi\,\sqrt{\beta}}
\ .
\ee
\par
Upon solving the above Eq.~\eqref{mT}, and considering the ``physical''
branch (which reproduces the Hawking behavior for $m\gg\mpl$),
one obtains
\be
T
&\!\!=\!\!&
\frac{1}{4\,\pi \beta}
\left(
m-\sqrt{m^2-\beta\,\mpl^2}
\right)
\\
&\!\!=\!\!&
\frac{1-\sqrt{1-\beta}}{4\,\pi\beta}
\left(
\mpl-
\frac{m-\mpl}{\sqrt{1-\beta}}
\right)
+
{\mathcal O}\left[(m-\mpl)^2\right]
\ ,
\nonumber
\ee
for $0<\beta<1$, where we expanded around $m\simeq\mpl$.
It is interesting to note that such expression for $T$ is still meaningful also for $\beta < 0$.
These possibilities hint at a lattice micro structure of the space-time,
and have been explored, e.g.~in polymer quantization, and in world crystal
physics~\cite{WC}.
\par
Recalling now that the emission rate can be written as
\be
\frac{\d m}{\d t}
=
-\frac{8\, \pi^3\,m^2\,T^4}{15\,\mpl^5\,\lp}
\ ,
\label{SB}
\ee
we obtain the decay rate
\be
-\frac{\d m}{\d t}
\simeq
\alpha \,\frac{m^2}{\mpl\,\lp}
+
{\mathcal O}(m-\mpl)
\ ,
\label{GUPdm}
\ee
for $T\simeq T_{\rm p} = \mpl$ (or $m\simeq \mpl$),
where $4\cdot10^{-5} < \alpha < 7\cdot10^{-4}$ when $0<\beta<1$.
\par
It is perhaps questionable that objects with a mass of the order of $\mpl$ can
be described by the usual thermodynamical arguments, which stem from a
(semi-)classical picture of black holes.
However, the horizon wave-function for a particle was precisely conceived to describe
this quantum regime, and we can now assume that the probability the black hole decays
is given by the probability $P_{\rm T}$ that the particle can be found outside its own
horizon~\footnote{The subscript T is for tunnelling, which is reminiscent of the interpretation
of the Hawking emission as a tunnelling process through the horizon~\cite{PW}.
Note, however, that the horizon is fuzzy in our description and not a (backreacting)
classical surface.}.
Of course, if the mass $m\ll\mpl$, the horizon wave-function tells us the particle is
most likely not a black hole to begin with, so the above interpretation must be restricted
to $m\simeq \mpl$ (see again Fig.~\ref{2Psi}).
\par
We first define
\be
P_>(r>\rh)
=
P_{\rm S}(r>\rh)\,P_{\rm H}(\rh)
\ ,
\label{PrmoreH}
\ee
where now
\be
P_{\rm S}(r>\rh)
=
4\,\pi\,\int_{\rh}^\infty
|\psi_{\rm S}(r)|^2\,r^2\,\d r
\ .
\ee
Upon integrating the above probability over all values of $\rh$,
we then obtain (since $m\sim\ell^{-1}$)
\be
P_{\rm T}(m)
=
1-P_{\rm BH}(m)
\ ,
\ee
and, expanding (\ref{Pbh2}) for $m\simeq\mpl$,
\be
P_{\rm T}(m)
\simeq
a-b\,\frac{m-\mpl}{\mpl}
\ ,
\ee
where $a\simeq 0.14$ and $b\simeq 0.65$ are positive constants of order one.
The amount of the particle's energy that can be found outside the horizon could thus
be estimated by
\be
\Delta m
\simeq
m\,P_{\rm T}
\simeq
a\,m
+
{\mathcal O}(m-\mpl)
\ .
\ee
At the same time, from the time-energy uncertainty relation
\be
\Delta E\,\Delta t
\simeq
\mpl\,\lp
\ ,
\ee
one obtains the typical emission time
\be
\Delta t
\simeq
\frac{\lp^2}{\Delta \rh}
\simeq
\ell
\ ,
\ee
where we used (\ref{hoop}) and (\ref{DRH}).
Putting the two pieces together, we then find that a near Planck size black hole
would emit according to
\be
-\frac{\Delta m}{\Delta t}
&\!\!\simeq\!\!&
a\,\frac{m}{\ell}
+
{\mathcal O}(m-\mpl)
\nonumber
\\
&\!\!\simeq\!\!&
a\,\frac{m^2}{\mpl\,\lp}
+
{\mathcal O}(m-\mpl)
\ ,
\label{qbh}
\ee
in functional agreement with the prediction from the GUP given in Eq.~\eqref{GUPdm}.
\par
It is now important to remark that there is a fairly large numerical discrepancy between
the numerical coefficients in Eq.~\eqref{GUPdm} and those in Eq.~\eqref{qbh}. 
For once, this disparity can perhaps be traced back to the fact that, with Eq.~\eqref{SB},
we are applying the canonical formalism to a Planck mass particle, which is not completely sensible,
since the particle/black hole should be in quasi equilibrium with its radiation for thermodynamical
arguments to hold.
The horizon wave-function, instead, knows nothing of the thermodynamics, and should have
therefore a more general validity.
However, we must point out that the above description of black hole evaporation relies on a
totally static representation of the Quantum Mechanical particle, and is
therefore to be viewed as a first attempt at modelling the decay of a quantum
black hole in the present picture.
A more accurate account of the microscopic structure of quantum black holes is indeed likely
to change the details (see, e.g.~Refs.~\cite{gomez,QHBH}), but the fact that 
this simple treatment leads to results similar to those following from the GUP is
already intriguing, and suggestive that an even more accurate Quantum Mechanical
description should be possible.
Finally, let us mention that in this Planckian regime, regardless of the microscopic model,
it would certainly be more appropriate to use the microcanonical formalism~\cite{CH}
(based on energy conservation, a property not entailed by the GUP).
Future work will be devoted to refine the calculation in these all of these directions.
\section{Conclusions and outlook}
\label{conc}
We have here introduced a horizon wave-function as a tool that allows us to effectively
describe the emergence of a horizon in a localised Quantum Mechanical system.
For the simple case of a spherically symmetric massive particle, the horizon wave-function
already supports the existence of a minimum black hole mass, without assuming {\em a priori\/}
the existence of a minimum (fundamental) length~\cite{BNSminimallength,piero}~\footnote{The
existence of this mass threshold may have phenomenological implications in models with extra spatial
dimensions~\cite{add,rs}, where the fundamental (gravitational) length
corresponds to energy scales potentially as low as a few TeV's.}.
Moreover, it does so in a genuinely Quantum Mechanical fashion, since it produces
a negligible probability that a particle with mass much smaller than $\mpl$ is a black hole,
rather than giving a sharp value for the particle mass above which the transition from particle
to black hole occurs.
Further, the description of black holes that the horizon wave-function entails was shown
to be compatible with GUPs, since it yields the same kind of uncertainty relation in phase
space, and a similar decay rate for Planck size objects.
\par
The results presented here should be however viewed as preliminary, as the notion
of a horizon wave-function requires a thorough generalisation before it can be effectively
employed to analyse more interesting physical problems.
We already mentioned in the Introduction that it is of particular conceptual interest
to study the possibility of black hole production in high-energy collisions~\cite{Kanti:2008eq,fischler}.
Let us here recall that, along these lines, Dvali and co-workers~\cite{dvali} recently went on to
conjecture that the high-energy limit of all physically relevant Quantum Field Theories
involves the formation of a (semi)classical state (to wit, black hole formation
for gravity), which should automatically suppress trans-Planckian quantum fluctuations.
This idea extends the concept of a GUP to include gravity, as was considered, for example in
Refs.~\cite{scardigli,BNSminimallength} and implies that the mass of microscopic black holes
must be quantized, and admit a minimum value~\cite{dvalimin} (for more general cases,
see also Ref.~\cite{visser}).
Beside the conceptual relevance for the inclusion of gravity in a description of
all forces of nature, there is also the potential phenomenological relevance of
quantum mechanical effects during the formation of trapping horizons and
black holes of astrophysical size.
\par
In fact, one should not forget that the basic building blocks of matter remain the
Standard Model particles, and that at such extreme energy regimes quantum effects
should not be easily overlooked.
All of the above conjectures would therefore be conspicuously substantiated if we could
understand the extremely complex dynamics of colliding Standard Model particles,
including the effect of the gravitational interaction, around the Planck scale~\cite{veneziano,fischler}.
To this purpose, the definition of the horizon wave-function for simple spherical systems
must be generalised to describe particle collisions and the inclusion of angular momentum
in the initial and final configurations~\cite{InProgress}.
It appears hard to complete such steps without a more detailed model of ``quantum black holes'',
in order to define the Hilbert space of the horizon wave-function.
One could, for example, incorporate the conjecture of Refs.~\cite{gomez} and~\cite{QHBH},
and describe the matter sourcing the black hole geometry as a condensate at the phase
transition.
\acknowledgements
R.C.~would like to thank O.~Micu and B.~Harms for useful comments.
R.C.~is supported by the I.N.F.N.~grant BO11.
F.S. would like to thank Misao Sasaki, for warm hospitality at Yukawa Institute, Kyoto,
where some early stages of this work were conceived.
\end{document}